\documentclass[MSNbibl,nameyear,dvips]{arxstspdf}
\usepackage{graphicx}
\usepackage{flushend}
\usepackage{stfloats}

\volume{26}
\issue{4}
\pubyear{2011}
\firstpage{584}
\lastpage{597}
\doi{10.1214/11-STS356}



\begin{document}
\begin{frontmatter}

\title{Multiple Testing for Exploratory Research\thanksref{T1}}
\relateddois{T1}{Discussed in \relateddoi{d}{10.1214/11-STS356A},
\relateddoi{d}{10.1214/11-STS356B} and \relateddoi{d}{10.1214/11-STS356C};
rejoinder at \relateddoi{r}{10.1214/11-STS356REJ}.}

\runtitle{Multiple Testing for Exploratory Research}

\begin{aug}
\author{\fnms{Jelle J.} \snm{Goeman}\corref{}\ead[label=e1]{j.j.goeman@lumc.nl}}
\and
\author{\fnms{Aldo} \snm{Solari}\ead[label=e2]{aldo.solari@unimib.it}}
\runauthor{J. J. Goeman and A. Solari}

\affiliation{Leiden University Medical Center and University of Milano-Bicocca}

\address{Jelle J. Goeman is Associate Professor, Department of Medical
Statistics and Bioinformatics (S5-P), Leiden University Medical Center,
P.O. Box 9600, 2300 RC Leiden, The Netherlands \printead{e1}.
Aldo Solari is Assistant Professor, Department of Statistics, University of Milano-Bicocca, Italy \printead{e2}.}
\end{aug}

\begin{abstract}
Motivated by the practice of exploratory research, we formulate an
approach to multiple testing that reverses the conventional roles of
the user and the multiple testing procedure. Traditionally, the user
chooses the error criterion, and the procedure the resulting rejected
set. Instead, we propose to let the user choose the rejected set
freely, and to let the multiple testing procedure return a confidence
statement on the number of false rejections incurred. In our approach,
such confidence statements are simultaneous for all choices of the
rejected set, so that post hoc selection of the rejected set does not
compromise their validity. The proposed reversal of roles requires
nothing more than a~review of the familiar closed testing procedure,
but with a focus on the non-consonant rejections that this procedure
makes. We suggest several shortcuts to avoid the computational problems
associated with closed testing.
\end{abstract}


\vspace*{-4pt}
\end{frontmatter}

\section{Introduction}

Central to the practice of statistics is the distinction between
exploratory and confirmatory data analysis, and the interplay between
the two. Exploratory data analysis suggests and formulates hypotheses,
which can subsequently be rigorously\break tested by confirmatory data
analysis. The two types of data analysis require very different methods
(Tu\-key, \citeyear{Tukey1980}): where\vadjust{\goodbreak} confirmatory data
analysis is structured and rigorous, exploratory data analysis can be
open-minded and speculative.

Hypothesis testing and strict Type I error control are traditionally
part of the realm of confirmatory data analysis, and, by implication,
so are multiple testing procedures. However, multiple hypothesis
testing is increasingly finding its way into exploratory data analysis.
In genomics research, for example, typical experiments test thousands
of hypotheses corresponding to as many molecular markers. Although
somewhat structured, such experi-\break ments~should be viewed as exploratory
rather than as confirmatory. The collection of tested hypotheses is
usually not selected on the basis of any theory, but because it is
convenient and exhaustive. The rejected hypotheses are generally not
meant to be reported as end results, but are to be followed up by
independent validation experiments.

Despite the exploratory nature of these experiments, researchers do
feel a need for multiple hypothesis testing methods and, in fact,
routinely apply them. The main reason for this is that researchers want
to protect themselves from following up on too many false leads and
doing too many unsuccessful validation experiments.\vadjust{\goodbreak} Most multiple
testing methods, however, have been designed for confirmatory data
analysis and are ill-suited for the specific requirements of
exploratory research.

Before we come to the main argument of this paper, we would like to set
the scene by sketching the requirements for an inferential procedure
for exploratory research. Imagine the situation that we are exploring a
large, but finite number of candidate hypotheses, indiscriminately
selected. Rather than rigorously proving the validity of some or all of
these hypotheses, as in confirmatory analysis, we want to select a
number of promising hypotheses for further probing in a next phase of
validation. The open-minded nature of exploratory research can be
described by three characteristics: exploratory research is
\textit{mild}, \textit{flexible} and \textit{post hoc}. We explain these
three terms below, contrasting them with the more familiar
characteristics of confirmatory research.

An inferential procedure is \textit{mild} if it allows some false
positives among the selected hypotheses. This is the most obvious
characteristic of exploratory research. Mildness is reasonable because
false positives are expected to be detected and removed in later
validation experiments. Confirmatory research, in contrast, being the
final phase of the research cycle, is not mild but strict.

An inferential procedure is \textit{flexible} if it does not prescribe to
the researcher which precise hypotheses to select or not to select. For
example, if the procedure ranks the hypotheses from most to least
promising, but the researcher detects a common\break theme in the hypotheses
ranked second, third and fourth, he or she can choose to follow up on
these three hypotheses and disregard the hypothesis that ranked first.
In fact, the researcher may also choose to follow up on the hypothesis
that ranked last, if that fits the same theme. Such freedom, ``picking
and choosing,'' is an important part of the hypothe\-sis-generating
aspect of exploratory research. In confirmatory research, in contrast,
selection of an interesting and coherent collection of hypotheses has
been done prior to the experiment, so that flexible selection is not
necessary.

Finally, an inferential procedure is \textit{post hoc} if it allows all
choices that are inherent to the procedure to be made after seeing the
data. Specifically, how mild the procedure should be, and which precise
set of hypotheses to select does not have to be chosen beforehand, but
may be chosen on the basis of the data. This is probably the most
distinguishing feature of exploratory research. The decision which
inferences, and how many, to follow up is often based on a mixture of
considerations; these considerations are usually not purely
statistical, and are often difficult to make explicit. In contrast, in
pure confirmatory research all choices regarding the testing procedure
have to be set in stone before data collection.

An ideal multiple hypothesis testing procedure for exploratory research
should sanction a mild, post hoc and flexible approach. Multiple
testing procedures generally do not fulfil these criteria. The main
present distinction is between multiple testing methods based on the
familywise error (FWER), and variants, and methods based on the false
discovery rate (FDR), and variants of that.

FWER-based methods control the probability of making any false
rejection at a prespecified rate. These are the archetypical methods
for confirmatory analysis. Such methods are clearly not mild, and they
are not post hoc, as all data analysis decisions have to be made before
seeing the data. They can be argued to be flexible in a limited sense:
it is possible to refrain from rejecting some of the rejected
hypotheses without violating control of the familywise error, but it is
not possible to reject any hypotheses that were not selected by the
procedure. A variant of familywise error, $k$-FWER, has been formulated
that controls the probability of making at least $k\geq1$ false
rejections (Romano and Wolf, \citeyear{Romano2007}). Depending
on $k$, methods with this error rate are mild and are flexible in the
same limited way as FWER itself is. Still, $k$-FWER-based methods have
so far only attracted theoretical interest as in these methods value of
$k$ may not be chosen post hoc, and nobody knows how to choose $k$ a
priori in an applied setting. A~recent permutation method of
Meinshausen (\citeyear{Meinshausen2006}) can be seen as a method that controls $k$-FWER
simultaneously for all values of~$k$, and consequently allows post hoc
selection of~$k$. This method is mild, post hoc, and quite flexible,
although it does not allow a fully arbitrary selection of the set of
rejected hypotheses.

False Discovery Rate (Benjamini and Hochberg, \citeyear{Benjamini1995}) methods control the expected proportion of
falsely rejected hypotheses among the rejected hypotheses. Such methods
are not very well suited for traditional confirmatory research and take
a step toward exploratory research. FDR-based methods are certainly
mild compared to FWER-based methods. However, they are not post hoc, as
the set of rejected hypotheses is completely determined after setting
the FDR threshold. Moreover, FDR-based methods are not flexible: as
shown by Finner and Roters (\citeyear{Finner2001}),\vadjust{\goodbreak} and illustrated in a practical example by
Marenne et al. (\citeyear{Marenne2009}), selecting a subset from the hypotheses that the
FDR-controlling procedure rejects may increase the false discovery rate
above the prespecified level, just like, of course, selecting a
superset can. Many variants of FDR have been proposed
(e.g., Storey, \citeyear{Storey2002}; Efron et al., \citeyear{Efron2001};
Van Der Laan, Dudoit and Pollard, \citeyear{Laan2004}), but none of these has
the desired three characteristics of the ideal multiple testing
procedure for exploratory inference. Methods have been formulated for
selective inference (Benjamini and Yekutieli, \citeyear{Benjamini2005}),
but these still do not allow the full
flexibility of exploratory selection.

In this paper we present an approach to multiple testing that does
allow mild, flexible and post hoc inference. By the nature of the
requirements of being flexible and post hoc, such a procedure cannot
prescribe what hypotheses to reject, but can only advise. This reverses
the traditional roles of the user and the procedure in multiple
testing. Rather than, as in FWER- or FDR-based methods, to let the user
choose the quality criterion, and to let the procedure return the
collection of rejected hypotheses, the user chooses the collection of
rejected hypotheses freely, and the multiple testing procedure returns
the associated quality criterion. In our view, the task of a multiple
testing procedure in the exploratory context is not to dictate what to
reject, but to quantify the risk taken, in terms of the potential
number of false rejections, of following up on any specific set of
hypotheses, chosen freely.

This reversal of roles can be achieved while avoiding the pitfall of
proposing yet another variant of FWER or FDR; it can be done simply by
combining the familiar concept of the confidence set, the discrete
version of the confidence interval, with the well-known closed testing
procedure (Marcus, Peritz and Gabriel, \citeyear{Marcus1976}), widely
recognized as a fundamental principle of multiple testing. What we will
show is that the closed testing procedure can be used to construct
exact simultaneous confidence sets for the number of false rejections
incurred when rejecting any specific set of hypotheses, measuring the
risk of following up on this particular set of hypotheses. Because the
confidence sets are simultaneous over all possible sets of rejected
hypotheses, the user is free to optimize, making the procedure valid
even under post hoc selection of the rejected set.\looseness=1

The approach we propose is constrained by the requirement that the
number of hypotheses potentially\vadjust{\goodbreak} to be followed up is finite and that
these hypotheses can be listed a priori. While this requirement rules
out the most open-minded and unstructured applications of exploratory
research, many exploratory problems are structured enough to fit the
framework.

Our proposed procedure has strong links to $k$-FWER methods. In fact,
the constructed confidence sets can be seen as controlling the
$k$-FWER, but simultaneously for all values of $k$, thus sanctioning
post hoc selection of $k$ and removing the requirement of selecting $k$
a priori, which traditionally plagues $k$-FWER-based methods. Through
this, our method links to the approach of Meinshausen (\citeyear{Meinshausen2006}); we
come back to this link in Section~\ref{section simes}.

Another interesting link is with methods that have appeared in recent
years for estimating $\pi_0$, the number of true hypotheses among the
collection of all hypotheses
(Schweder and Spj{\o}tvoll, \citeyear{Schweder1982}; Benjamini and Hochberg, \citeyear{Benjamini2000};
Langaas, Lindqvist and Ferkingstad, \citeyear{Langaas2005}; Meinshausen and B{\"u}hlmann, \citeyear{Meinshausen2005};
Jin and Cai, \citeyear{Jin2007}). The procedure outlined in this paper automatically gives a
confidence set for the quantity~$\pi_0$, because the collection of all
hypotheses is one of the possible sets of rejected hypotheses that the
user can choose to follow up, and the number of false rejections in
that set is exactly~$\pi_0$.

The outline of this paper is as follows. In the next section, we review
the closed testing procedure and the role of the concept of consonance
in that procedure. We argue that non-consonant closed testing
procedures have been underrated, and illustrate the type of additional
inference that is possible from a non-consonant closed testing procedure,
but typically neglected, before we argue how these additional
inferences can be used to construct a confidence set. Section
\ref{stepwise} applies the approach to selection of variables in a
multiple regression model. Section~\ref{shortcuts} explores
computational issues related to closed testing procedures and proposes
situations in which shortcuts can be found. Finally,
Section~\ref{estimation} looks at estimation of the number of correctly
rejected hypotheses. Software to perform the methods described in this
paper is available in the \textit{cherry} package, downloadable from
CRAN.

\section{Non-consonant Closed Testing} \label{theory}

The closed testing procedure (Marcus, Peritz and Gabriel,
\citeyear{Marcus1976}) is well known as a cornerstone of familywise
error control. In this section we show how closed testing may also be
used to construct confidence sets for the number of falsely rejected
hypotheses.\vadjust{\goodbreak}

\begin{figure}

\includegraphics{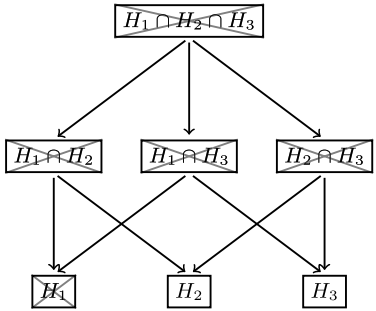}

\caption{Intersection hypotheses formed by elementary hypotheses $H_1$,
$H_2$ and $H_3$. Rejected hypotheses have been marked with a cross. The
rejection of $H_2\cap H_3$ is a non-consonant rejection.}
\label{example}
\end{figure}

First we introduce some notation. Let $H_1, \ldots, H_n$ be the
collection of hypotheses of interest, the \textit{elementary hypotheses},
out of which we want to select hypotheses to follow up. Some of these
hypotheses are true; let $T \subseteq \{1, \ldots, n\}$ denote the
unknown indices of true hypotheses. To use a closed testing procedure,
we must consider not only the elementary hypotheses, but also all
intersection hypotheses of the form $H_I = \bigcap_{i\in I} H_i$, where
$I \subseteq \{1, \ldots, n\}$, $I \neq \varnothing$. Figure
\ref{example} illustrates the intersection hypotheses formed by three
hypotheses $H_1$, $H_2$ and $H_3$ in the form of a graph, with arrows
denoting subset relationships (ignore the crosses for now).

An intersection hypothesis $H_I$ is true whenever all $H_i$, $i\in I$,
are true, that is, whenever $I \subseteq T$. Let the closure
$\mathcal{C}$ be the collection of all nonempty subsets of the index
set $\{1, \ldots, n\}$. Each element of $\mathcal{C}$ corresponds to an
intersection hypothesis, some of which are true. Let $\mathcal{T} = \{
I \in \mathcal{C}\dvtx I \subseteq T\}$ be the subsets corresponding to
true intersection hypotheses. The collection $\mathcal{C}$ also
contains singleton sets. Noting that we can equate $H_i = H_{\{i\}}$,
let $\mathcal{H} = \{I \in \mathcal{C}\dvtx \#I=1\}$ be the subsets
corresponding to the elementary hypotheses.

The closed testing procedure works as follows. It requires
$\alpha$-level tests for every intersection hypothesis $H_I$,
$I\in\mathcal{C}$, which are called the \textit{local tests}. Applying
these local tests, let $\mathcal{U} \subseteq \mathcal{C}$ be the
collection of subsets $U \in \mathcal{C}$ for which the test rejects
the hypotheses $H_U$. The collection $\mathcal{U}$ represents the raw
rejections uncorrected for multiple testing. Based on these raw
rejections, the closed testing procedure rejects every $I \in
\mathcal{C}$ for which $J \in \mathcal{U}$ for every $J \supseteq I$.
Denote the collection of all such $I$ by $\mathcal{X}$. It was shown
very elegantly by Marcus, Peritz and Gabriel (\citeyear{Marcus1976}) that with this rejected set the
closed testing procedure strongly\vadjust{\goodbreak} controls the familywise error for all
hypotheses $H_I$, $I \in \mathcal{C}$, at level $\alpha$. They showed
that the event $E = \{H_T \notin \mathcal{U}\}$, which happens with
probability at least $1-\alpha$, implies that $\mathcal{X} \cap
\mathcal{T} = \varnothing$.

In the example of Figure~\ref{example}, suppose that the hypotheses
rejected by the local tests are the ones marked with a cross. In this
example $H_1$ is rejected by the closed testing procedure because the
four hypotheses $H_1$, $H_1 \cap H_2$, $H_1 \cap H_3$ and $H_1 \cap H_2
\cap H_3$ are all rejected by their local test. In fact, in the example
of Figure~\ref{example} we have $\mathcal{X} = \mathcal{U}$, because
each hypothesis rejected by the local test has all its ancestors in the
graph of Figure~\ref{example} rejected.

When using the closed testing procedure for familywise error control,
the intersection hypotheses are generally constructed for the benefit
of the procedure, but are not of genuine interest by themselves. The
reported result of the procedure is therefore usually not the
collection $\mathcal{X}$, but only $\mathcal{X} \cap \mathcal{H}$. From
the perspective of familywise error control, a rejection  $I \in
\mathcal{X}$ for which there is no $J \in \mathcal{X} \cap \mathcal{H}$
with $J \subset I$ is a wasted rejection. Such a rejection was not
instrumental in facilitating a rejection of interest; if that rejection
had not occurred, the same rejected set $\mathcal{X} \cap \mathcal{H}$
of elementary hypotheses would have resulted from the procedure. This
consideration has led to a quest for consonant closed testing
procedures. A closed testing procedure is \textit{consonant} if the local
tests for every $I \in \mathcal{C}$ are chosen in such a way that
rejection of $I$ implies rejection of at least one $J \in \mathcal{H}$.
It is easily shown that for every closed testing procedure there is a
consonant procedure that rejects at least as much in $\mathcal{X} \cap
\mathcal{H}$. Moving from a non-consonant to a consonant procedure may
often lead to a gain in power on the elementary hypotheses. From a
familywise error perspective, consonance is, therefore, a desirable
property, and non-consonant procedures are best avoided
(Bittman et al., \citeyear{Bittman2009}).

However, once we are interested in milder inference than a familywise
error-based one, the premise that only rejection of the elementary
hypotheses $H_1,\allowbreak \ldots, H_n$ is of interest should be dropped, and
non-consonant closed testing procedures need not be\break avoided. We
illustrate this with the simple example of Figure~\ref{example}, which
will immediately serve as a small showcase of the point of view on
multiple testing we propose in this paper. Here, the only one of the
elementary hypotheses that has been rejected is $H_1$. Of the
intersection hypotheses we see three ``consonant'' rejections, namely
$H_1 \cap H_2 \cap H_3$, $H_1 \cap H_2$ and $H_1 \cap H_3$, which have
all facilitated rejection of the elemental hypothesis $H_1$. We also
see one ``non-consonant''\vadjust{\goodbreak} rejection, $H_2 \cap H_3$. A~familywise error
perspective would dictate rejection of $H_1$ and nothing else. An
exploratory perspective, however, on the same data would lay the choice
what and how many hypotheses to reject with the user. An obstinate user
could, for example, choose not to reject $H_1$, but to reject $H_2$ and
$H_3$. What can we say about the risk incurred by such a user in terms
of the number of false rejections?

In general, the number of false rejections made when rejecting the
hypotheses $H_i$, $i\in R$, is equal to $\tau(R) = \#(T \cap R)$, the
number of true null hypotheses among~$R$. For a given set $R$, this
quantity is just a function of the model parameters, for which we can
find estimates and confidence intervals just like for any other
function of the model parameters. The confidence interval takes the
form of a confidence set, because $\tau(R)$ only takes discrete values.
We come to the issue of estimation later, and first construct such a
confidence set.

To construct a confidence set, define
\[
\mathcal{C}_R = \{I \in \mathcal{C}\dvtx I \subseteq R\},
\]
the collection of all intersection hypotheses involving only rejected
hypotheses, and let
\[
t_\alpha(R) = \max \{\#I\dvtx I \in \mathcal{C}_R, I \notin \mathcal{X}\},
\]
taking $t_\alpha(R) = 0$ if $\mathcal{C}_R \subseteq \mathcal{X}$. The
quantity $t_\alpha(R)$ is the size of the largest subset of $R$ for
which the corresponding intersection hypothesis is not rejected by the
closed testing procedure. We claim that the set
\begin{equation} \label{conf set}
\{0, \ldots, t_\alpha(R)\}
\end{equation}
is a $(1-\alpha)$-confidence set of the parameter $\tau(R)$.

To prove the coverage of this set, remember that if the event $E$ has
happened, then all rejections that the closed testing procedure has
made are correct. Given that $E$ has happened, the value of $\tau(R)$
cannot be greater than the value of $t_\alpha(R)$, because otherwise a
true intersection hypothesis would have been rejected, which is
inconsistent with the definition of $E$. Consequently, $\tau(R) \in
\{0, \ldots, t_\alpha(R)\}$ with probability at least
$\mathrm{P}(E)\!=\!1\!-\!\alpha$, which makes $\{0, \ldots,\allowbreak t_\alpha(R)\}$ a
$(1-\alpha)$-confidence set for $\tau(R)$.

The confidence set (\ref{conf set}) is always one-sided, never
providing a nontrivial lower bound for $\tau(R)$. The reason for this
is that the confidence set originates from a procedure that is focused
on rejecting, not on accepting null hypotheses. Furthermore, for many
applications the null hypotheses are point hypotheses, of which it can
never be proved that they are true. In these cases, no procedure can
produce a confidence interval with a nontrivial lower bound, and the
upper bound is the only bound of real interest.

Often interest is in quantifying not the number of true hypotheses in
$R$, but the number of false hypotheses $\phi(R) = \#R-\tau(R)$. A
confidence set for $\phi(R)$ follows from (\ref{conf set}) immediately
as
\[
\{f_\alpha(R), \ldots, \#R\},
\]
where $f_\alpha(R) = \#R - t_\alpha(R)$. Confidence sets for other
quantities that depend only on $\tau(R)$ and $\#R$, such as the false
discovery proportion $\tau(R)/\#R$, may be derived in a similar way.

Returning to the example of Figure~\ref{example} with choice of a
rejected set, $R = \{2, 3\}$, we have a realized value of
$t_\alpha(R)=1$. We conclude that $\{0,1\}$ is a
$(1-\alpha)$-confidence set for the number of false rejections incurred
when rejecting $H_2$ and $H_3$. Even though neither $H_2$ or $H_3$ was
rejected by the closed testing procedure, when rejecting both $H_2$ and
$H_3$ the user can be confident of making at least one correct
rejection. The choice of $R = \{2, 3\}$ is only one of many possible
rejection choices that the user can make. For each alternative choice,
a confidence set can be made in the same way as for $R = \{2, 3\}$.
These confidence sets, and the corresponding confidence sets for
$\phi(R)$, are given in Table~\ref{table example}.

\begin{table}
\caption{Confidence sets for the numbers of incorrect rejections
$\tau(R)$ and correct rejections $\phi(R)$ incurred with various
choices of the rejected set, based on the closed testing result of
Figure \protect\ref{example}} \label{table example}
\begin{tabular*}{\columnwidth}{@{\extracolsep{\fill}}lcc@{}}
\hline
$\bolds{R}$ & \textbf{Confidence set for} $\bolds{\tau}\bolds{(R)}$ & \textbf{Confidence set for}
$\bolds{\phi}\bolds{(R)}$\\
\hline
  $\{1\}$ & \{0\} & \{1\}\\
  $\{2\}$ & \{0, 1\} & \{0, 1\} \\
  $\{3\}$ & \{0, 1\} & \{0, 1\} \\
  $\{1, 2\}$ & \{0, 1\} & \{1, 2\} \\
  $\{1, 3\}$ & \{0, 1\} & \{1, 2\} \\
  $\{2, 3\}$ & \{0, 1\} & \{1, 2\} \\
  $\{1,2,3\}$ & \{0, 1\} & \{2, 3\} \\
  \hline
\end{tabular*}
\vspace*{3pt}
\end{table}

The important thing to note about confidence sets of the form
(\ref{conf set}) is that they are simultaneous confidence sets, which
all depend on exactly the same event $E$ for their coverage. Because
these confidence sets are simultaneous, the user can review all these
confidence sets, and select the rejected set $R$ that he or she likes
best, while still keeping correct $1-\alpha$ coverage of the selected
confidence set: under the event~$E$, all confidence sets cover the true
parame\-ter simultaneously, and therefore, under the same event~$E$, the
selected confidence set covers the true parame\-ter. Consequently, the
selected confidence set has coverage of at least $\mathrm{P}(E) =
1-\alpha$. The simulta\-neity of the sets makes their coverage robust
against post hoc selection.

In the specific case of Table~\ref{table example}, the user might
choose to follow up on all three hypotheses, which would give him or
her confidence in at least two discoveries of a false null hypothesis.
On the other hand, if sufficient funds are available for only two
validation experiments, the user may choose to follow up on any two
hypotheses, any pair giving confidence of obtaining at most one false
positive.

Contrary to the application of closed testing for familywise error
control, in terms of confidence sets non-consonant rejections do
improve the results obtained from the procedure. Without the rejection
of $H_2 \cap H_3$ in Figure~\ref{example} the confidence sets for $R =
\{2, 3\}$ and for $R = \{1, 2, 3\}$ would have been larger than the
ones given in Table~\ref{table example}. From the definition of
consonance it follows immediately that the value of $t_\alpha(R)$ in a
consonant closed testing procedure is equal to the number of hypotheses
in $R$ that are not rejected by the closed testing procedure under a
familywise error regime. In non-consonant closed testing procedures,
the value of $t_\alpha(R)$ can be substantially smaller, as we shall
see in examples below.\looseness=1

Essentially, the example of Table~\ref{table example} summarizes the
confidence set approach to multiple testing. The user has unlimited
options in selecting what to reject, and may review all options and
their consequences in order to make his or her choice. This approach
fulfills all three criteria set for multiple testing in exploratory
research formulated in the introduction. The procedure is flexible,
because it does not prescribe any rejections but leaves the choice
which hypotheses to follow up completely in the hands of the user. The
procedure is mild, because it allows any number or proportion of false
rejections that the user desires. Furthermore, the procedure is post
hoc, because it allows the user to review the consequences, in terms of
the potential number of false rejections, of any choice of rejected
hypotheses before making a final choice, without compromising the
quality of the inferences obtained. Still, even with the lenience of
all these properties, the inferential statements resulting from the
procedure are absolutely classical and rigorous, requiring no new
definitions of error rates but only the classical concept of
simultaneous confidence sets.

\section{Example: Selecting Covariates in~Regression} \label{stepwise}

One area of statistics in which common practice is highly exploratory
and post hoc is the selection of covariates in a multiple regression.
Methods such as forward or backward selection, or their combination,
are typically used to select a model containing a subset of a set of
candidate covariates. Often, $p$-values that are reported for the
selected covariates completely ignore the selection process. The
confidence set method outlined in the previous section can be used in
this situation to set confidence limits to the number of selected
variables that is truly associated with the response variable.

As an example, consider the \textit{physical} dataset
(Lar\-ner, \citeyear{Larner1996}), in which 10 physical
measurements on 22 male subjects (length, and circumference of various
parts of the body) are used as covariates for modeling body mass. An
analysis based on a linear regression model with a forward--backward
algorithm selects the four covariates \textit{forearm}, \textit{waist},
\textit{height} and \textit{thigh} as the relevant variables. Table
\ref{physical} gives the $p$-values of the covariates in both the full
and the selected model. The reported $p$-values of the selected model
are known to be anti-conservative as they do not take the selection
into account. An important question to ask, therefore, is how many
truly relevant variables are, in fact, included in this selection. This
would give a measure of confidence for the selected set.

\begin{table}
\caption{Uncorrected $p$-values ($t$-test) for relevance of variables
in the full model and selected model} \label{physical}
\begin{tabular*}{\columnwidth}{@{\extracolsep{\fill}}lcc@{}}
\hline
\textbf{Covariate} & \textbf{Full model} & \textbf{Selected model} \\
\hline
(Intercept)& 0.036 &  0.000  \\
Forearm       & 0.061 &  0.000   \\
Biceps      & 0.755 &  --       \\
Chest      & 0.420 &  --        \\
Neck       & 0.518 &  --        \\
Shoulder   & 0.905 &  --       \\
Waist      & 0.000 &  0.000      \\
Height     & 0.033 &  0.005      \\
Calf       & 0.303 &  --        \\
Thigh      & 0.351 &  0.036      \\
Head       & 0.105 &  --    \\
\hline
\end{tabular*}
\end{table}

Following the strategy outlined in the previous section, we construct a
linear regression model with an intercept and 10 regression
coefficients $\beta_1,\ldots, \beta_{10}$, and define the elementary
hypotheses $H_i$, $i=1,\ldots,\allowbreak10$, to be the hypotheses that the
corresponding regression coefficient $\beta_i=0$. Next we construct all
1,023 intersection hypotheses $H_I$, $I\in \mathcal{C}$, each of which
corresponds to the hypothesis that $\beta_j=0$ for all $j \in I$. As
the local tests we choose the $F$-test of the corresponding null model
against the saturated model, tested at level $\alpha = 0.05$.

The closed testing procedure rejects 626 out of the 1023 hypotheses,
among which there is one elementary hypothesis: \textit{waist}. Several
non-consonant rejections have occurred. We can summarize these by
finding the \textit{defining} rejections, that is, the rejections $I \in
\mathcal{X}$ which have no rejected subset $J \subset I$, $J \in
\mathcal{X}$. For this dataset, these defining rejections are the
following seven sets:\vspace*{6pt}

\begin{tabular}{l}
\{\textit{waist}\} \\
\{\textit{forearm,     neck,     shoulder, height}\} \\
\{\textit{forearm,     biceps,    shoulder, calf}\} \\
\{\textit{forearm,     shoulder, height,   calf}\} \\
\{\textit{forearm,     biceps,    chest,    neck,     shoulder, thigh}\} \\
\{\textit{forearm,     shoulder, height,   thigh}\} \\
\{\textit{forearm,  calf,  thigh}\}
\end{tabular}\vspace*{6pt}

As each of these sets corresponds to a rejected intersection
hypothesis, we can conclude with 95\% confidence that each of the seven
sets contains at least one truly relevant covariate. It is tempting to
say that, beside \textit{waist}, \textit{forearm} must be relevant, since
it is included in all defining sets except the first. This is not
warranted, however, as the sets are also consistent with alternative
truths, such as that both \textit{shoulder} and \textit{thigh} are relevant
variables. What we can conclude, is that if we select, for example, the
set
\begin{equation} \label{optimalset}
R =\{ \mbox{waist, forearm,  calf,  thigh}\}
\end{equation}
we have selected at least two relevant variables. Furthermore, we can
also directly conclude that \textit{waist} is a relevant variable.

Coming back to the set $R = \{\mbox{waist},\ \mbox{forearm},\break \mbox{height},\ \mbox{thigh}\}$
selected by the forward--backward procedure, we can find all 15
intersection hypotheses of the four hypotheses in $R$ and check whether
they were rejected by the closed testing procedure. We find that
$R\in\mathcal{X}$, but $\mbox{\{forearm,  height,  thigh\}} \notin
\mathcal{X}$, so that $t_\alpha(R) = 3$. Therefore, we can say with
confidence that the selected set $R$ contains one truly relevant
hypothesis, but not that it contains more than one. From this result,
it is clear that the $p$-values given for the selected model in Table
\ref{physical} are highly untrustworthy.

To find out how many of the original 10 hypotheses are relevant, we
take $R$ to be the full set of 10 hypotheses, and we calculate
$t_\alpha(R)=8$ for this set. Apparently, we can conclude that there
are at least two covariates among these 10 that are determinants of
mass. The smallest set that contains at least two relevant covariates
is the set (\ref{optimalset}).

It should be noted that the set selected by variable selection
procedure should not generally be expected to be optimal from a
confidence set perspective, because the perspectives of the two
procedures are quite different. This is best illustrated by thinking of
a dataset in which there are two covariates which are both highly
correlated with each other, and with the response. Variable selection
algorithms will always choose one of the two variables, disregarding
the second one as superfluous given the first. The confidence set
approach, however, will emphasize the uncertainty of the choice between
the two variables, and will not reject any intersection hypothesis that
involves only one of the two covariates. To have confidence that at
least one truly relevant covariate is included, both of the highly
correlated variables must be selected. This reflects a difference in
emphasis between the two approaches: variable selection selects optimal
sets, whereas the confidence set approach quantifies the uncertainty
inherent in the selection process.

It is interesting to investigate the price of post hoc selection
relative to a priori selection. It is immediate from the procedure that
reducing the tested set of hypotheses to a set $R$ a priori is at least
as powerful as, and likely more powerful than, testing a larger set and
selecting the same set $R$ post hoc. Post hoc selection will generally
result in wider confidence sets than a priori selection: this is the
price to be paid for the risk of overfit caused by post hoc selection.
In the example this price is surprisingly small. If the set $R =
\{\mbox{waist, forearm,  height,  thigh}\}$ would have been defined a priori as
the set of hypotheses of interest, treating the remaining covariates'
regression coefficients as nuisance parameters, the confidence set of
$\phi(R)$ improves from $\{1,2,3,4\}$ to $\{2,3,4\}$. The confidence
set for $\phi(R)$ for the set~$R$ defined in (\ref{optimalset}) does
not change.\vspace*{-1pt}

\section{Shortcuts} \label{shortcuts}\vspace*{-1pt}

In its standard form, application of a closed testing procedure
requires $2^n-1$ tests to be performed. Smart bookkeeping can reduce
this number somewhat, especially if some intersection hypotheses high
up in the hierarchy turn out non-significant, because it can be used
that if $I \notin \mathcal{X}$, then immediately $J \notin \mathcal{X}$
for every $J \subset I$, which saves calculation of some of the tests.
Still, even with such tricks and with high computational power, the
closed testing procedure becomes computationally intractable in its
general form for a number of hypotheses around 20--30, depending on the
computational effort needed for each single test.

If a large number of hypotheses is to be investigated, it is,
therefore, convenient if the local tests can be chosen in such a way
that not all these tests need to be calculated. Methods for avoiding
calculation of some of the hypothesis tests in the closed testing
procedure are known as \textit{shortcuts}. The literature on shortcuts in
the closed testing procedure has been focused mainly on consonant
procedures, and on finding the rejected individual hypotheses
(Grechanovsky and Hochberg, \citeyear{Grechanovsky1999jspi}; Zaykin et~al., \citeyear{Zaykin2002};
Hommel, Bretz and Maurer, \citeyear{Hommel2007};
Bittman et~al., \citeyear{Bittman2009};
Brannath and Bretz, \citeyear{Brannath2010jasa}). In this section, we loosely extend the concept of
shortcuts to non-consonant procedures, and discuss ways of finding
$t_\alpha(R)$ in a computationally easy way for specific choices of the
local test, namely those based on Fisher combinations, on Simes'
inequality, and on sums of normally distributed test statistics. We
also demonstrate how the permutation-based procedure of
Meinshausen (\citeyear{Meinshausen2006}) fits into the closed testing framework. Finally,
we touch upon the possible use of other procedures than closed testing
for constructing confidence sets.

\subsection{Fisher Combinations} \label{Fisher}

The case of independent null hypotheses deserves special attention as a
benchmark, because several important multiple testing methods
(Benjamini and Hochberg, \citeyear{Benjamini1995}; Efron et~al., \citeyear{Efron2001};
Storey, \citeyear{Storey2002}) have been initially
formulated for independent hypotheses only. Independent tests are
relatively rare in practical applications.

One highly suitable choice for the local tests in the independent case
is Fisher's combination method. It requires only the $p$-values $p_1,
\ldots, p_n$ of the tests of the elemental hypotheses $H_1, \ldots,
H_n$, and rejects an intersection hypothesis corresponding to $I \in
\mathcal{C}$ whenever
\[
-2 \sum_{i\in I} \log(p_i) \geq g_{\#I},
\]
where $g_r$ is the $(1-\alpha)$-quantile of a $\chi^2$-distribution
with $2r$ degrees of freedom.\vadjust{\goodbreak} This test is a valid $\alpha$-level test
of the hypothesis $H_I$ if the $p$-values $p_i$, $i\in I$, are
independent. Note that the requirement of being a valid local test only
refers to intersection null hypotheses that are true, so that there is
no requirement of independence among $p$-values of false null
hypotheses, nor even between $p$-values of true and false null
hypotheses.

\begin{table}
\caption{Adverse events data, taken from Herson (\protect\citeyear{Herson2009}), sorted to
increasing $p$-values and with a typo corrected} \label{Herson}
\begin{tabular*}{\columnwidth}{@{\extracolsep{\fill}}lc@{}}
\hline
\textbf{Adverse event} & $\bolds{p}$\textbf{-value}\\
\hline
Anemia & 0.02 \\
Myocardial infarct & 0.03 \\
Diarrhea & 0.04 \\
Nausea and vomiting & 0.04 \\
Stomatitis & 0.08 \\
Skin rash & 0.10 \\
Dehydration & 0.12 \\
Shortness of breath & 0.18 \\
Renal failure & 0.20 \\
Fever & 0.23 \\
Blurred vision & 0.26 \\
Nose bleed & 0.28 \\
Anorexia & 0.30 \\
Bronchitis & 0.31 \\
Wheezing & 0.40 \\
Headache & 0.50 \\
\hline\\
\end{tabular*}
\end{table}

Fisher's method is highly non-consonant, as sum test often are.
Moreover, the simple structure of the local tests allows easy shortcuts
to be formulated. For any $s < \# R$, we have that $t_\alpha(R) \leq s$
if
\[
u(R, s+1) \geq \max_{0\leq j\leq M} \{ g_{s+j+1} - u(\bar R,j) \},
\]
where $u(I,k)$ is the sum of the $k$ smallest values of $-2\log(p_i)$
with $i \in I$, $\bar R$ is the complement of $R$, and~$M$ is the
number of values of $-2\log(p_k)$ in $\bar R$ smaller than the
$(s+1)$th largest value of $-2\log(p_k)$ in $R$. This shortcut, related
to the shortcut of Zaykin et al. (\citeyear{Zaykin2002}), allows calculation of
$t_\alpha(R)$ for any $R$ without exponentially many tests having to be
calculated. It is an example of a general method for finding shortcuts
for exchangeable tests, which we explain in Appendix~\ref{exchangeable}.

As an example, consider the following application in the realm of
adverse drug reactions. Consider the data in Table~\ref{Herson}, which
give raw $p$-values for null hypotheses concerning the presence of
adverse drug reactions reported for a certain drug. We assume the
hypotheses to be independent, although the validity of this assumption
can be disputed.

An analysis based on familywise error rate ({\v{S}}id{\'a}k, \citeyear{Sidak1967})
or false discovery rate (procedure of Benjamini and Hochberg, \citeyear{Benjamini1995})
results in no rejections for these data. However,
among the hypotheses with small $p$-values, the researcher notices
three hypotheses concerned with problems of the gastrointestinal tract:
diarrhea, nausea and vomiting, and stomatitis. The researcher may
hypothesize that the drug causes problems in this area, and may
consider following up on these three hypotheses. For this choice of $R$
we can calculate $f_\alpha(R)=1$ at $\alpha=0.05$, and we can conclude
that the drug in question has at least some adverse effect somewhere in
the gastrointestinal tract. The researcher can be confident that
following up on these three hypotheses will lead to at least one
potentially successful validation experiment.

\begin{figure}

\includegraphics{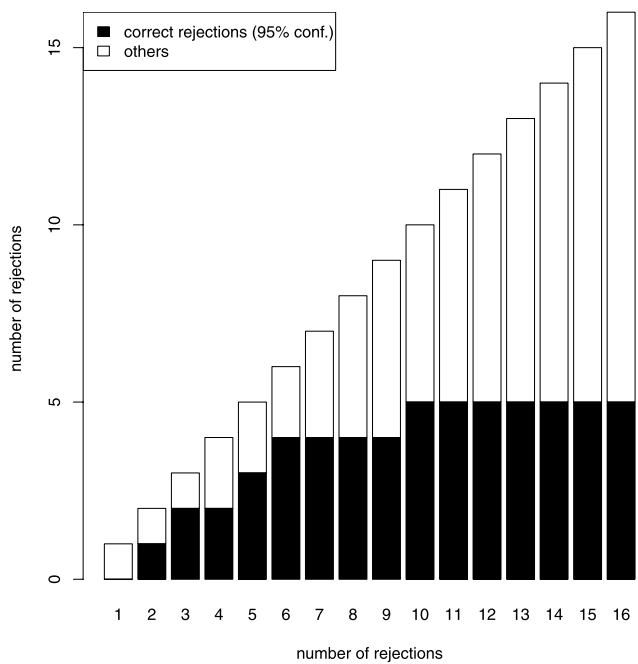}

  \caption{Number of correct rejections versus number of
  rejections for the data of Table \protect\ref{Herson}. The bars only
  give the lower bound of the 95\% confidence interval; the number of
  false null hypotheses (correct rejections) is likely to be larger
  than indicated.} \label{barplot}
\end{figure}

Alternatively, if the researcher wants to optimize the number of
correct rejections, he or she may simply wish to reject those
hypotheses that have the smallest $p$-values. In that case the only
choice the researcher has to make is the number of rejections, and a
plot such as Figure~\ref{barplot} may be made, which plots the lower
bound of the number of correct rejections $f_\alpha(R)$ against the
number of rejections $\#R$. Based on this plot, the researcher can
claim with 95\% confidence\vadjust{\goodbreak} that at least five adverse drug reactions
occur for this drug, and that these are found among the hypotheses with
the 10 smallest $p$-values. If the researcher does not have funds
available for 10 follow-up experiments, the researcher may want to
validate the top six, which gives confidence of finding at least four
false null hypotheses, or perhaps the top three, for confidence of
finding at least two false null hypotheses.

Figure~\ref{barplot} also illustrates the link between our proposed
approach and the $k$-FWER criterion. A user wishing to control $k$-FWER
can reject any set $R$ that has $t_\alpha(R) < k$, for example taking
$R$ as the set corresponding to the $i$ smallest $p$-values, choosing
$i$ as the largest value such that $t_\alpha(R) < k$ still holds. The
graph of Figure~\ref{barplot} simultaneously shows the numbers of
rejections allowed with $k = 1,2,3,4,\ldots,$ which are given by $i=
0,3,6,7,\ldots.$ A major advantage of our approach over traditional
$k$-FWER control approaches is that control is simultaneous over all
rejected sets, and therefore over all choices of $k$. The procedure
thus bridges the gap between weak FWER control, related to $n$-FWER,
and strong 1-FWER control. Furthermore, rather than choosing~$k$ in
advance, its value may be picked after seeing the data without
destroying the associated control property. The link between $k$-FWER
and our approach is not limited to local tests based on Fisher
combinations.

An interesting feature of using Fisher's method in combination with the
confidence set approach is that the method may prove the presence of
false null hypotheses even in the situation that no individual
$p$-value is smaller than $\alpha$. Consider the following \mbox{$p$-values},
taken from Huang and Hsu (\citeyear{Huang2007}):
\begin{eqnarray*}
&&p_1=0.051;\quad p_2=0.064;\\
 &&p_3=0.097;\quad p_4=0.108.
\end{eqnarray*}
Even though all $p$-values are non-significant individually, the
confidence set for $\phi(R)$ when rejecting the top two hypotheses is
$\{1,2\}$, when rejecting the top three hypotheses $\{2,3\}$, and when
rejecting all four hypotheses $\{2,3,4\}$. This indicates that even in
absence of any individually significant hypotheses we can make a
rigorous confidence statement that at least two out of the first three
hypotheses are false.

Fisher's method is highly non-consonant, and can be very powerful,
especially if there are many moderately small $p$-values. It is not
uniformly more powerful than other tests, however. Compared to
consonant local tests, such as Sidak's, Fisher's method tends to have\vadjust{\goodbreak}
smaller values of $t_\alpha(R)$ for large rejected sets $R$ due to its
large number of non-conso\-nant~rejections, but Sidak's method often has
more sets~$R$ which have $t_\alpha(R)=0$, due to a higher number of
consonant rejections.

\subsection{Simes Type Local Tests and Permutations} \label{section simes}

A different type of local test with potentially non-consonant
rejections is a type of test that rejects a~hypothesis $H_I$, $I \in
\mathcal{C}$, whenever
\begin{equation} \label{simes style}
p^I_{(i)} \leq c^{\#I}_i
\end{equation}
for at least\vspace*{1pt} one $1 \leq i \leq \#I$, where $p^I_{(i)}$ is the $i$th
smallest among the $p$-values $\{p_j\}_{j\in I}$ of the elementary
hypotheses with indices in $I$, and $c_i^m$, \mbox{$1\leq m \leq n$}, $1 \leq
i \leq m$, are appropriately chosen critical values. Without loss of
generality we can take $c^m_i \leq c_j^m$ if $i \leq j$.

We call local tests of the form (\ref{simes style}) \textit{Simes type}
local tests because if we choose
\begin{equation} \label{simes}
c_i^m = \frac{i\alpha}{m},
\end{equation}
the test based on (\ref{simes style}) is a valid $\alpha$-level test of
$H_I$ by Simes' (\citeyear{Simes1986}) inequality. Simes' inequality
holds whenever $p$-values of true null hypotheses are independent, but
also under more general conditions, as investigated by
Sarkar (\citeyear{Sarkar1998}). In particular it holds for $p$-values from
identically distributed, nonnegatively correlated test statistics.

A variant of Simes' inequality has been proposed by Hommel (\citeyear{Hommel1983}).
This variant uses critical values
\begin{equation} \label{simes hommel}
c_i^m = \frac{i\alpha}{K_m\cdot m},
\end{equation}
where $K_m = \sum_{v=1}^{m} v^{-1}$. Unlike the one based on Simes'
inequality, the local test defined by these critical values is of the
correct level $\alpha$ whatever the dependence structure of the
original $p$-values.

Local tests of the form (\ref{simes style}) do not generally allow
shortcuts for the calculation of $t_\alpha(R)$, but two useful
shortcuts are available if the critical values are chosen in such a way
that
\begin{equation} \label{simes condition shortcut}
c_i^l \leq c_i^m\quad \mbox{whenever } l \geq m.
\end{equation}
The first shortcut this condition allows is the general shortcut
described in Appendix~\ref{exchangeable}, the conditions of which are
fulfilled whenever (\ref{simes condition shortcut}) holds. The second
shortcut is even faster to calculate, but is less general: it holds for
rejected sets of the form $R = \{i \dvtx p_i \leq q\}$ only. Let $p_{(i)}$,
$i=1,\ldots,n$, be short for $p_{(i)}^I$ with $I = \{1,\ldots,n\}$. For
$R$ of the form mentioned, we have the shortcut
\begin{equation} \label{simes shortcut}
f_\alpha(R) > \max\{S_r\dvtx 1\leq r\leq \#R\},
\end{equation}
where $S_r = \max\{s\geq 0\dvtx p_{(r)} \leq c^n_{r-s}\}$. The value of
$S_r$ can be interpreted as the number of more stringent critical
values $c^n_{r-1}, \ldots, c^n_{1}$ by which the $p$-value~$p_{(r)}$
overshoots its mark $c^n_r$. The number of false hypotheses is larger
than the greatest such overshoot of the ordered $p$-values in the set
$R$. The shortcut~(\ref{simes shortcut}) is useful for making plots
such as the one in Figure~\ref{barplot}. We prove this shortcut in
Appendix~\ref{shortcut 2}. A~slightly more powerful variant of the
shortcut (\ref{simes shortcut}) is available if we have (\ref{simes condition shortcut}) and
\begin{equation} \label{simes condition shortcut strong}
c_i^m \geq c_{i-w}^{m-w}\quad \mbox{for every $1 \leq w < i$.}
\end{equation}
In this case, we have the same shortcut as (\ref{simes shortcut}), but
with $S_r = \max\{s\geq 0\dvtx p_{(r)} \leq c^{n-s}_{r-s}\}$. The proof of
this statement is analogous to the proof for (\ref{simes shortcut}) and
is also given in Appendix~\ref{shortcut 2}. It is easy but tedious to
show that the Simes critical values (\ref{simes}) and (\ref{simes
hommel}) conform to (\ref{simes condition shortcut}) and that the
critical values (\ref{simes}) also conform to (\ref{simes
condition shortcut strong}), so that the shortcuts may be used for
these choices of the local test (see also Benjamini and Heller, \citeyear{Benjamini2008}).

As a side note, we remark that the critical values~(\ref{simes}) and
(\ref{simes hommel}) are the same as the critical values used in the
false discovery rate controlling algorithms of Benjamini and Hochberg (\citeyear{Benjamini1995}) and
Benjamini and Yekutieli (\citeyear{Benjamini2001}), respectively. The correspondence between the
critical values creates a connection between the corresponding methods.
The set that has been rejected by the false discovery rate algorithm
always has $f_\alpha(R) > 0$ based on the closed testing procedure with
the corresponding local test. Note that the assumptions underlying each
local test and its corresponding false discovery rate algorithm are
very similar. For the example data of Section~\ref{Fisher}, the Simes
local test leads to no rejections, which is consistent with finding no
rejections with the procedure of Benjamini and Hochberg (\citeyear{Benjamini1995}).

Permutation testing can be a powerful tool to take into account the
joint distribution of the $p$-values. Useful shortcuts in a closed
testing procedure with permutation-based local tests can be constructed
from the work of Meinshausen and B\"uhlmann (\citeyear{Meinshausen2005})
and Meinshausen (\citeyear{Meinshausen2006}).
These authors describe\break a~permu\-tation-based way to find critical values
$k_i$, $i=1,\ldots,n$, such that the probability under the complete null
hypothesis that $p_{(i)} \leq k_i$ for at least one $i$ is bounded by~$\alpha$. The same method may in principle also be used to find
corresponding permutation critical values $k_i^I$ for every
intersection hypothesis~$H_I$, $I \in \mathcal{C}$, and therefore a
local test for every intersection hypothesis $H_I$; a closed testing
procedure can be made on the basis of these tests. However, unless the
number of hypotheses is limited, this will be extremely time-consuming,
and it would lead to a~closed testing procedure for which shortcuts are
not available. A way out of this dilemma can be found by remarking
that, by construction of the permutation critical values, we have
$k_i^I \geq k_i$ for every~$i$ and~$I$. Therefore, a valid, though
conservative, local test may be constructed by simply using a procedure
of the form (\ref{simes style}) with $c_i^m = k_i$ for every $1\leq m
\leq n$. This local test fulfils the condition (\ref{simes condition
shortcut}) and therefore admits shortcuts. With this choice of a local
test, the confidence set approach to multiple testing can be used for
every collection of test statistics for which permutation is possible,
opening up the possibility to use permutation-based closed testing in
genomics research.

Meinshausen (\citeyear{Meinshausen2006}) constructed simultaneous confidence bands for
the number of falsely rejected hypotheses for rejected sets of the form
$R = \{i \dvtx p_i \leq q\}$, similar to Figure~\ref{barplot}, based on the
permutation critical values $k_1,\ldots, k_n$ he found. Even though
Meinshausen did not use closed testing, these confidence bands are
identical to the confidence bounds that would be obtained when using
the local tests (\ref{simes style}) with $c_i^m = k_i$ in combination
with the shortcut (\ref{simes shortcut}). By exploiting the shortcut of
Appendix~\ref{exchangeable} rather than this simpler shortcut, it
becomes possible to extend Meinshausen's method to be able to find
confidence bounds for $\tau(R)$ for sets $R$ not of the form $R = \{i
\dvtx
p_i \leq q\}$. Alternatively, for a very small number of tests, the
full permutation-based closed testing procedure may be used, which
could be more powerful.

\subsection{Normally Distributed Test Statistics} \label{normal}

Workable local tests may also be constructed on the basis of normally
distributed scores. Consider the situation that we have scores $z_1,
\ldots, z_n$ for each hypothesis $H_1, \ldots, H_n$, respectively,
which are standard normally distributed if their respective null
hypothesis is true, and we would reject $H_i$ one-sidedly when $z_i$ is
large. This situation occurs quite frequently in practice, at least
asymptotically, for example if we do many one-sided binomial $z$-tests.
A~sensible choice for a test statistic for a local test is $Z_I =
\sum_{i\in I} z_i$. Consider first the case in which the scores\vadjust{\goodbreak} of true
null hypotheses are independent. In that case $Z_I$ is normally
distributed with mean 0 and variance $\#I$, and we may reject $H_I$
whenever $Z_I \geq \sqrt{\#I} \cdot \Phi(1-\alpha)$, where $\Phi$ is
the standard normal distribution function. If the scores are not
independent but only jointly normally distributed, we have the
following, more conservative result. In that case~$Z_I$ is normally
distributed with mean 0 but unknown variance. Let $\Sigma$ be the
correlation matrix of $\{z_i\}_{i\in I}$, then the variance of $Z_I$ is
given by $\mathbf{1}^\mathrm{T}\Sigma\mathbf{1}$, where
$\mathbf{1}$ is a vector of ones of length $\#I$. This variance is
bounded by $\# I$ times the largest eigenvalue of $\Sigma$, and
therefore by $(\# I)^2$. It follows that for $\alpha \leq 1/2$, we may
reject $H_I$ whenever $Z_I \geq \#I \cdot \Phi(1-\alpha)$. This type of
test was used by Van De Wiel, Berkhof and Van Wieringen (\citeyear{Wiel2009}).

Both tests are exchangeable and lead to easy shortcuts in the sense of
Appendix~\ref{exchangeable}. In practice, the test for the
non-independent case can be highly conservative if used for small
values of $\alpha$, unless the scores are strongly positively
correlated. One case to note, however, is the case that $\alpha=1/2$,
when the critical value is 0 for both the independent and the general
situation, negating the conservativeness of the latter. This situation
is relevant for the method of Section~\ref{estimation}.

\subsection{Other Types of Shortcuts}

Shortcuts of the form described in the appendices can only be used
within a restricted class of local tests that is calculated as an
exchangeable function of per-hypothesis statistics. Other types of
shortcuts may be devised for other classes of local tests in the
future.

A very different way to construct confidence intervals of $\tau(R)$
while avoiding calculation of the complete closed testing procedure is
to use a different multiple testing procedure that still allows
non-con\-sonant rejection of some intersection hypotheses. Examples of
such procedures are the tree-based testing procedure of
Meinshausen (\citeyear{Meinshausen2008}), recently improved by Goeman and Solari (\citeyear{Goeman2010}), the
focus level procedure of Goeman and Mansmann (\citeyear{Goeman2008}), and the gatekeeping method
of Edwards and Madsen (\citeyear{Edwards2007}). These procedures allow familywise error
inference on a collection of hypotheses comprising the elementary
hypotheses and a selection from the $2^n-1$ intersection hypotheses,
and may produce non-consonant rejections on these intersection
hypotheses. The results of these procedures may be used as a basis for
constructing confidence intervals in the same way as the results of the
closed testing procedure were used in Section~\ref{theory}.\vadjust{\goodbreak}

\section{Estimation} \label{estimation}

In addition to the confidence interval, it can sometimes be informative
to have a point estimate of the number of true null hypotheses among a
set of interest. Estimation of the number of true null hypotheses has
been a subject of recent interest in the context of genomic data
analysis, and several authors
(Schweder and Spj{\o}tvoll, \citeyear{Schweder1982}; Benjamini and Hochberg, \citeyear{Benjamini2000};
Langaas, Lindqvist and Ferkingstad, \citeyear{Langaas2005};
Meinshausen and B{\"u}hlmann, \citeyear{Meinshausen2005}; Jin and Cai, \citeyear{Jin2007})
have proposed methods for estimating $\tau(R)$, although for $R =
\{1,\ldots,n\}$ only. The quantity $\tau(R)$ for $R = \{1,\ldots,n\}$
is commonly referred to as $\pi_0$.

The confidence intervals of the previous sections are easily adapted to
produce a point estimate of~$\tau(R)$ for any set $R$. We propose to
use the value $t_{1/2}(R)$ as an estimate, the upper bound of the
confidence interval, calculated at the significance level $\alpha =
1/2$. This estimate can be seen as a conservative median estimate of
the true quantity $\tau(R)$: by the properties of $t_\alpha(R)$ derived
in the previous sections, $t_{1/2}(R)$ exceeds the value of $\tau(R)$
with a probability that is bounded above by $1/2$. Furthermore, this
property holds simultaneously for all $R$ by the simultaneity of the
confidence interval on which it is based, which makes the defining
property of the estimate robust against selection of $R$.

The estimate can be used to get an impression where the ``midpoint'' of
the confidence interval is. Applying the procedure  to the physical
dataset of Section~\ref{stepwise} at $\alpha=1/2$, we find the
following defining rejections:\vspace*{6pt}

\begin{tabular}{l}
\{\textit{waist}\} \\
\{\textit{forearm}\} \\
\{\textit{height}\} \\
\{\textit{chest, calf, thigh}\} \\
\{\textit{neck, calf, thigh}\} \\
\{\textit{thigh, head}\}
\end{tabular}\vspace*{6pt}

\noindent The estimated number of true null hypotheses among all 10
hypotheses, for which the 95\% confidence set was $\{0, \ldots, 8\}$,
is calculated as 6, which points to an estimated number of four
relevant variables in the regression. For the set $R =
\{\mbox{waist},\ \mbox{forearm},\ \mbox{height},\break \mbox{thigh}\}$ of four variables selected by the stepwise
procedure, the number of truly relevant variables is estimated at 3
(the 95\% confidence set for this quantity was $\{1,2,3,4\}$). In
contrast, the set (\ref{optimalset}) which included with 95\%
confidence at least two relevant variables, also contains an estimated
number of two relevant variables. The smallest rejected set that
contains an estimated number of four relevant variables is the set $R
= \{\mbox{waist},\ \mbox{forearm},\ \mbox{height},\ \mbox{thigh},\break \mbox{head}\}$. We can report this
optimized set without fear of overfit, because the property that the
number of truly relevant covariates is overestimated with probability
at most $1/2$ holds simultaneously for all rejected sets.

In the adverse event data of Section~\ref{Fisher} the number of true
null hypotheses among the 16 hypotheses is estimated at two using a
Fisher local test at $\alpha=1/2$. In this case, all rejections turn
out to be consonant: rejecting the 14 hypotheses with smallest
$p$-values leads to an estimated number of 0 false discoveries. If we
use Simes rather than Fisher for the local test, we even obtain an
estimated number of 0 true null hypotheses among all 16 hypotheses.

We warn against using the estimate of the number of falsely rejected
hypotheses by itself, without the associated confidence interval. To
see the danger of this, consider the simplest ``multiple testing
problem'' in which only a single null hypothesis is tested. The
estimation procedure of this section would estimate this hypothesis as
true whenever the $p$-value is greater than $1/2$, and as false whenever
it is smaller than or equal to $1/2$. This seems generally too lenient a
conclusion to be a viable strategy, although it may be useful in some
highly exploratory and risk-seeking settings. In these situations, the
special status of $\alpha=1/2$ in the shortcut of Section~\ref{normal}
may be of interest.

\section{Conclusion}

All exploratory research is essentially picking and choosing. From a
large number of potential hypotheses to follow up, the researcher
selects for further investigation those hypotheses or sets of
hypotheses that stand out in the researcher's eyes. This selection is
made in complete freedom. The notion that any statistical method would
dictate what the researcher should find interesting is contrary to the
spirit of exploratory research.

However, a well-known risk of picking and choosing is overfit,
``cherry-picking.'' Patterns that strike the researcher as relevant and
interesting may have arisen due to chance, and turn out to be false
positives in follow-up experiments. To protect a research\-er against too
many disappointments of this type, it is important to make a realistic
assessment of the risk taken when following up on a certain collection
of hypotheses.

In this paper, we have presented an approach to multiple testing that
is especially designed for the requirements of exploratory research,
and which reverses the way that multiple testing methods are typically
used. Rather than letting the user decide on the error rate, and the
procedure on the rejections, we let the user decide on the rejections,
and the procedure on the error rate. Our approach does not rely on the
definition of any new error rates, and has not even required the design
of a new algorithm. The approach uses the classical concept of the
simultaneous confidence set, together with the equally classical closed
testing procedure, although both in a novel way.

The end result of the procedure is a collection of confidence sets for
the number of falsely rejected hypotheses for all possible choices of
the rejected set. The most important property of these confidence sets
is that they are simultaneous. This simultaneity protects the user of
the procedure against overoptimism resulting from post hoc selection of
the rejected set, and removes many of the problems traditionally
associated with cherry-picking from a large set.

Finally, the approach is very general, and the limits on its useability
are mostly computational. The most important assumption we make is that
the number of hypotheses potentially to be followed up is finite, and
that these hypotheses may be enumerated before starting the experiment.
Aside from that, the ability of the closed testing procedure to work
with any choice of a local test makes that procedure very flexible.
Only if the number of hypotheses becomes large, computational issues
limit the choice of local tests to those for which shortcuts are
available. The shortcuts described in this paper already cover a wide
range of application areas. More and improved shortcuts are likely to
be found in the future.

\appendix

\section{Shortcuts for Exchangeable Local Tests} \label{exchangeable}
\renewcommand{\theequation}{\arabic{equation}}
\setcounter{equation}{8}

We present a fairly general method for constructing shortcuts in the
closed testing procedure which can be used for finding $t_\alpha(R)$
and are appropriate for the methods in Section~\ref{shortcuts}. This
shortcut delineates a class of local tests for which $t_\alpha(R)$ can
be calculated for any $R$ by calculating only~$n^2$, rather than~$2^n$
tests. We give the shortcut for a $p$-value-based method. The shortcut
for methods based on other scores (e.g., Section~\ref{normal}) is
completely analogous.\vadjust{\goodbreak}

Assume that the local test is exchangeable, that is, rejection of
$H_I$, $I \in \mathcal{C}$, only depends on the set $P_I = \{p_i\}_{i
\in I}$ of raw $p$-values, and not on the collection $I$ itself. Let
$\delta$ be the function that maps from a set of $p$-values $P$ to
rejection, $\delta(P)=1$ if the collection $P=P_I$ would lead to
rejection of $H_I$, and $\delta(P)=0$ otherwise. Further, suppose that
\begin{equation} \label{ass1}
\delta(\{p_1, \ldots, p_k\}) \geq \delta(\{q_1, \ldots, q_k\})
\end{equation}
whenever $p_1 \leq q_1, \ldots, p_k \leq q_k$, and that
\begin{equation} \label{ass2}
\delta(q \cup P) \geq \delta(P)
\end{equation}
whenever $q \leq \min(p\in P)$.

If these assumptions hold, it can be shown that for any $s < \# R$,
\begin{equation} \label{shortcut}
\hspace*{23pt}\delta(Q^R_{s+1} \cup \bar{Q}^R_j) = 1\quad \mbox{for every $j \in \{0,
\ldots, m_R\}$}\hspace*{-3pt}
\end{equation}
implies $t_\alpha(R) \leq s$. Here, $Q^R_{s+1}$ is the set of the $s+1$
largest $p$-values of hypotheses in $R$; $\bar{Q}^R_j$ is the set of
the $j$ largest $p$-values of hypotheses not in $R$, and $m_R$ is the
number of $p$-values not in $R$ that are larger than the smallest
$p$-value in $Q^R_{s+1}$.

To show this, note that by assumptions (\ref{ass1}) and~(\ref{ass2}),
equation (\ref{shortcut}) implies that
\[
\delta(Q^R_{s+1} \cup P_I) = 1
\]
for every $I \in \mathcal{C}$, and that therefore, by assumption~(\ref{ass1}),
\[
\delta(P_J \cup P_I) = 1
\]
for every $I \in \mathcal{C}$ and for every $J \subseteq R$ for which
$\# J = s+1$. Consequently, $J \in \mathcal{X}$ for every $J \subseteq
R$ for which $\# J = s+1$, so that $t_\alpha(R) \leq s$ by definition.

\section{Shortcuts for Simes-Type Local Tests} \label{shortcut 2}

Next, we prove the shortcut (\ref{simes shortcut}) for Simes-type local
tests. Let $R = \{i \dvtx p_i \leq q\}$ be a rejected set, and assume that
condition (\ref{simes condition shortcut}) holds.

First, let $r = \#R$, and remark that $p_{(r)} \leq c^n_{r-s}$, for
some $s \geq 0$, implies that
\begin{equation} \label{help proof simes shortcut}
f_\alpha(R) > s.
\end{equation}
To see why this is true, choose any $K \subseteq R$ with $\#K \geq r -
s$ and any $J \supseteq K$. Remark that $p_{(r)} \leq c^n_{r-s}$
implies that
\[
p_{(r-s)}^J \leq p_{(r-s)}^K \leq p_{(r)} \leq c^n_{r-s} \leq
c_{r-s}^{\#J}.
\]
Consequently, $K \in \mathcal{X}$ for every $K \subseteq R$ with $\#K
\geq r - s$, so that $t_\alpha(R) < r-s$, and (\ref{help proof simes
shortcut}) follows. To obtain the final statement (\ref{simes
shortcut}), remark that $f_\alpha(R) \geq f_\alpha(S)$ for every $R
\supseteq S$, and apply the bound (\ref{help proof simes shortcut}) on
all $S \subset R$ of the form specified.\vadjust{\goodbreak}

Analogously, if (\ref{simes condition shortcut}) and (\ref{simes condition shortcut strong}) both holds, choose
$K$ and $J$ as above, and let $\tilde s = \#(R\setminus J) \leq s$.
Then $p_{(r)} \leq c^{n-s}_{r-s}$ implies that
\[
p_{(r-\tilde{s})}^J \leq p_{(r)} \leq c^{n-s}_{r-s} \leq
c^{n-\tilde{s}}_{r-\tilde{s}} \leq c_{r-\tilde{s}}^{\#J},
\]
using (\ref{simes condition shortcut}) in the last inequality, noting that $\#J \leq n-\tilde{s}$ and that
(\ref{simes condition shortcut strong}) implies (\ref{simes condition
shortcut}). From this result (\ref{help proof simes shortcut}) and
(\ref{simes shortcut}) follow as above.


\end{document}